# Identity statuses in upper-division physics students

Paul W Irving . Eleanor C Sayre

**Abstract** We use the theories of identity statuses and communities of practice to describe three different case studies of students finding their paths through undergraduate physics and developing a physics subject-specific identity. Each case study demonstrates a unique path that reinforces the link between the theories of communities of practice and identity statuses. The case studies also illustrate how students progress and regress in their commitment to their subject-specific identities and their professional identities. The progression/regression is dependent on their willingness to explore different aspects of a physics professional identity and their availability to carry out such exploration. Identity status and future identity crises can manifest in students' behavior in the classroom. Allowing students to engage in more legitimate practices of the physics community, especially in the form of undergraduate research, helps students to explore their opportunities and inform the level of commitment they wish to make to physics.

## 

The development of a professional identity is a fundamental part of a higher education student's development, especially for students studying a subject that leads to professional status. An appropriate subject-specific identity is a strong influence on students' persistence in a discipline (Pierrakos et al. 2009). There is a strong relationship between the development of a professional, subject-specific identity and participation in a related community (Barton and Yang 2000); in fact, professional identity and community participation are inextricably and symbiotically linked (Tinto 1997). A key component of understanding subject-specific identity in higher education in general is understanding the choice to pursue a particular subject in the first place and why a student chooses to continue studying that subject. In the United States, choosing one's major or majors is a decision that many students make independently. For a large number of these students, it is the first substantial long-term planning decision they make autonomously (Galotti et al. 2006). We must understand how this decision relates to one's identity and the factors that influence and keep this decision in flux.

In this study, we gain insight into subject-specific choices in the context of upper-division physics students. We follow the trajectory of three students with differing paths through undergraduate physics. By combining the theories of identity status (Marcia 1966) and

P.W. Irving (✉)
Department of Physics, Michigan State University, East Lansing, MI 48824
e-mail: pwirving@gmail.com

E.C. Sayre (✉)
Department of Physics, Kansas State University, 116 Cardwell Hall, Manhattan, KS 66506-2601, USA
e-mail: esayre@ksu.edu

communities of practice (Lave and Wenger 1991), we can investigate how students' choices within and emotional connection (positive/negative affect) to the subject of physics change over time. Examining students' choices allows for an exploration of the connection between identity status and a student's position within their trajectory towards central membership in a community of practice. Identity-focused interviews and naturalistic observations of students interacting with their chosen subject in the classroom allow us to focus on these aspects and help provide a more complete picture of how students develop a subject-specific identity as a physicist.

**Identity**

Identity can and should be examined from multiple perspectives (Irving and Sayre 2013). The Communities of Practice framework (Lave and Wenger 1991) is one of the most prominent frameworks used to examine how people develop a professional identity. Communities of practice have three key characteristics: the individuals within a community form a group, either co-located or distributed (Coakes and Clark 2005); the group has common goals or shared enterprise (Wenger 1998); the group shares and develops knowledge focused on a common practice (Barab et al. 2002), including sharing mutually defined practices, beliefs, values, and history (Irving and Sayre 2014).

Within the context of the communities of practice framework, learning is conceived as a trajectory in which learners move from being legitimate peripheral participants to core participants of the community of practice in question (Barab et al. 2002). In the context of physics, this corresponds to students participating in legitimate peripheral practices of the community of practicing physics. Applying this framework to the college setting requires the interpretation of activities such as taking a quantum mechanics class or doing undergraduate research as legitimate peripheral practices of the community of practicing physicists. From this perspective, professors act as central participants of the community of practicing physicists, guiding peripheral participants on a trajectory to the core. As the students continue on this trajectory they gain more experience, skills, and knowledge which result in community cohesion and a sense of co-responsibility between participants (Wenger 1998) which is another important aspect of a community of practice.

The application of the communities of practice framework to the university context is further complicated by the fact that individuals participate in several overlapping communities of practice. A physics student involved in a research group might also be the goalie on a sports team, for example. The same research group might be part of a larger collaboration and, at the same time, members of the research group are also members of the physics department. Because one individual participates in several overlapping communities, it is important to study how 'more expansive networks' (Bonnar 2007) affect individuals' participation. Active participants in different communities of practice have opportunities to learn the knowledge, rituals, and histories valued within each community (Hsu and Roth 2010). Inasmuch as the different communities overlap, knowledge and practices learned in one community affect practices in another (Aschbacher et al. 2010). Conversely, when communities have different values, individual members may have difficulty importing practices from one community to another (Aikenhead 1996).

The focus in this study is on three case studies of physics students who navigate physics classes as undergraduates. In the past, we (Irving and Sayre, 2014) have examined three different communities of practice of different sizes concerned with physics students: physics classroom communities of practice, undergraduate physics community of practice, and the community of practicing physicists. The three students in this study are all members of these three different communities of practice (although perhaps not necessarily cognizant of their membership in each). In the case of classroom communities, they are in truth members of several each semester. It is within this context that a student develops a subject-specific identity in physics. A subject-specific identity does not refer to full central membership in the community of practicing physicists; instead, it is part of their trajectory to becoming a central member of the community of practicing physicists. It is important to remember that there are multiple communities of practice contributing to the development of a subject-specific identity and that the community of practice of undergraduate physics students is more closely aligned to developing a subject-specific identity (as opposed to a professional identity). In the case of students who are double or triple majoring, they also develop another subject-specific identity through membership to the same types of communities but in another discipline. There is overlap between the legitimate peripheral practices of the community of practice of undergraduate physics students and the community of practice of practicing physicists but there are also significant differences as illustrated in figure 1. Attending class and doing well on exams are legitimate practices of the community of practice of undergraduate physics students and also legitimate practices of the community of practicing physicists. On the other hand, conducting research, writing grant proposals, and attending conferences are more likely to be more central legitimate practices of the community of practicing physicists.

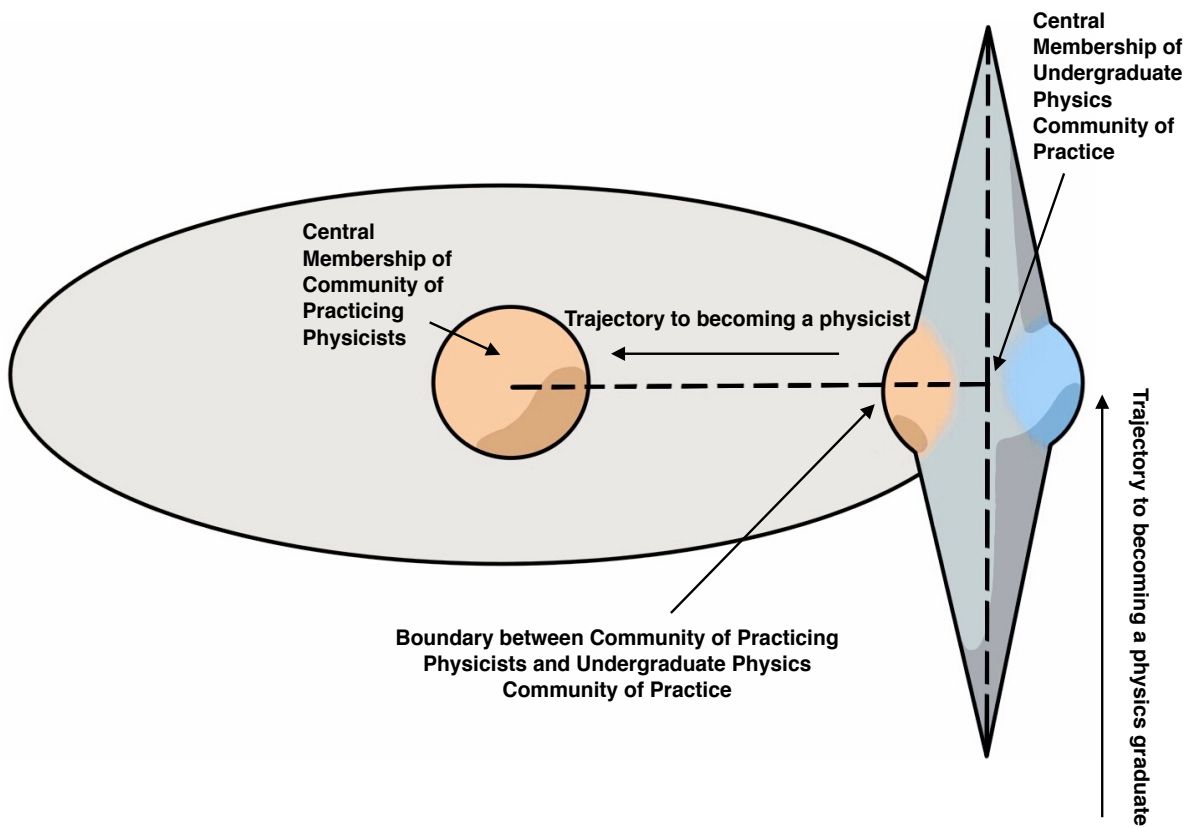

**Figure 1:** Connected communities of practice and trajectories

The case study students in this paper aspire to join several communities. They have already made the decision (whether it was a conscious decision or not) to be peripheral members of the community of practicing physicists. At the start of this research study, they are at different points in their trajectories to becoming core members of the community of practicing physicists. This study focuses on the important decision points (crisis points) where the students decide whether to choose to continue on this trajectory, to join other communities of practice, or to divert from this trajectory completely and call a halt to their progression towards core membership.

**Identity statuses**

"Identity status" is another prominent theory in identity research which allows the examination of these crisis points. Compared to the community of practice theory, which is a group orientated theory, the theory of identity statuses facilitates a focus on the individual.

The term "identity statuses" is used to describe the approaches individuals take in forming an identity (Marcia 1966). James Marcia developed the identity status paradigm to operationally define Eriksonian (Erikson 1968) identity formation (Berzonsky et al. 2013). Eriksonian identity development theory focuses on the psychosocial "identity crisis" (Busseri et al. 2011) in which adolescents test out their ideological and interpersonal domains. Therefore, identity statuses are typically focused on late adolescents and on determining their status through an examination along the dimensions of *exploration* and *commitment* to possible identities. *Exploration* is the questioning and deliberation of identities that are available for an adolescent to pursue. *Commitment* involves making a relatively firm choice about an identity domain and engaging in significant practices that are geared toward the implementation of that choice (Crocetti et al. 2013).

From this perspective, identity formation is seen as a number of identity crises. A crisis is a point in an adolescent's development where identity-relevant decisions need to be made (Germeijs et al. 2012). Commitments to an identity are then demonstrated by engaging in significant practices relevant to that identity. Marcia identified four identity statuses: *moratorium*, *identity achieved*, *foreclosed identity,* and *identity diffusion* (1966). Typically these are considered to be split into two groupings: those that have a high degree of commitment (*identity achieved* and *foreclosed identity*) and those that have a low degree of commitment (*moratorium* and *identity diffusion*).

The adolescents in the *identity achieved* status are highly committed to their identity decisions. A source of their high level of commitment is the transition to *identity achieved* status typically following a period of self-examination (Berzonsky et al. 2013). These individuals have a very firm focus on what they would like to do with their lives and their trajectory often remains undisturbed even when encountering external influences and naturally occurring obstacles. The strength of their commitment to their identity is a result of their previous explorations of the alternatives available to them (Luyckx et al. 2005).

Adolescents classified as having a *moratorium* identity status engage in the process of reflection and exploration. They have limited commitments to an identity during this process of reflection and are often conflicted between several identities which they could make commitments to. There are two likely outcomes of the process of reflection. The first, that the individual moves on from a moratorium status once a stable life commitment has been chosen (Adams et al. 2006) or the second, that the individual remains paralyzed in their indecision about which identity to subscribe to and end up in a continuous reflective process (Kroger and Marcia 2011).

Adolescents in the *foreclosed identity* status have made commitments to certain identities but with a relatively small amount of exploration (Crocetti et al. 2013). Individuals in the *foreclosed identity* can display just as much commitment to their identity as those individuals in an *identity achieved* status. However, further examination of their identity will reveal a reluctance to explore alternative identities in a genuine fashion. An individual in a *foreclosed identity* status is likely to choose an identity and career based on circumstantial reasons (Luyckx et al. 2010) and is more likely to become disenfranchised with their choices when influenced by obstacles and external influences.

Finally, the *identity diffusion* status refers to adolescents who have not engaged in either committing to or exploring alternative identities (Hirschi 2011). Individuals who are in the *identity diffusion* status do not engage with exploring their identity nor do they tend to make definite commitments (Kroger and Marcia 2011). In the context of this study and subject-specific identities, it would be very unusual for upper-level physics students to be in the *identity diffusion* status as they have already made major commitments to the subject by declaring it as a minor or major and continuing to take classes in the subject. It is still possible in a cycle of identity development that a student might retreat into a diffusive status if they become unhappy with the subject they have chosen to study.

Students will cycle through the three identity statuses of *foreclosed identity*, *moratorium*, and *identity achieved* at least once if not several times during their undergraduate career. Although it is possible to classify the movement from one status to another as either progressive or regressive (for example moratorium to achievement would be labeled as a progressive movement) the identity statuses are not unidirectional and they do not form a sequence by which individuals develop. In fact individuals can go through several status switches both progressively and regressively repeatedly over a long period (Krettenauer 2005).

More recent research (Crocetti et al. 2013) has argued for the existence of two different moratorium identity statuses: *classical moratorium* and *searching moratorium*. The difference between the two moratorium identity statuses is the origin of the motivation to reflect and explore. Individuals occupying the classical moratorium are unhappy with their current commitments and wish to explore and evaluate alternatives that will result in new identity-related commitments. Alternatively, the searching moratorium status is characterized by individuals who are not completely satisfied with their commitments (as opposed to unhappy) and wish to revise the commitments they have already made within the security provided by their initial commitments.

In a similar manner, recent research has also argued for the inclusion of two separate diffusion statuses (Luyckx et al. 2010). The main argument for differing diffusion statuses is the amount of effort one is expending to gain an identity. Individuals in the *diffused diffusion* status would try to explore future life plans but often in an uncoordinated and non-constructive manner. Alternatively, individuals classified as being in the *carefree diffusion* status do not waste energy on exploring and instead are content with the postponement of obtaining an identity.

|                                                    |                                                   |
| :------------------------------------------------: | :-----------------------------------------------: |
| Identity Achieved<br><br>High Commitment<br>High Exploration | Moratorium<br><br>Low Commitment<br>High Exploration |
| Foreclosed Identity<br><br>High Commitment<br>Low Exploration | Identity Diffusion<br><br>Low Commitment<br>Low Exploration |

**Figure 2**: Breakdown of identity status and level of commitment and exploration associated with each.

Figure 2 is a representation of the different identity statuses that will be used in the analysis section of this paper to illustrate individuals' changing status as they progress through their undergraduate career.

Research that is inherently linked to identity status research is that of identity processing styles. Previous research indicates that identity style and status are strongly correlated (Schwartz et al. 2000). Identity processing styles refer to social cognitive strategies which an individual will use to deal with a number of personal problems, specifically in relation to the context of this paper, the construction of a subject-specific identity and the making of personal/identity-related decisions (Berzonsky 2003). Previous research has identified three identity processing styles: *diffuse-avoidant, normative,* and *informational* (Adams et al. 2006).

The *diffuse-avoidant* style is characterized by procrastination, with adolescents accessing this style trying to avoid dealing with identity conflicts or decisions and, therefore, being very unlikely to form strong commitments, which is typically associated with students who have an identity diffusion status (Berzonsky and Ferrari 2009). The *diffuse-avoidant* style is also typically considered to be driven by the demands of each situation or crisis encountered and is perceived to be an irrational processing style (Berzonsky 1990).

The *normative* style is passive in nature with individuals who tend to adopt this style appropriating the goals, standards, and expectations held by important others with whom an individual has a relationship to (Adams et al. 2006). The normative style is considered to be automatic in nature resulting in a decreased expenditure of mental resources. However, it also tends to place a bias and distortion of the information being processed during a crisis and on the decision made to resolve the crisis (Berzonsky et al. 2013). Previous research has indicated a relationship between the *normative* processing style and the *foreclosed identity* status.

The *informational style* is firmly related to self-reflection. Individuals who access the *informational* style tend to be interested in learning new things about themselves and will change self-beliefs based on feedback. Previous research relates the *informational* style with an *achieved* or *moratorium* identity status (Berzonsky 2011). This focus on self-reflection makes the *informational* style relatable to the ideas of metacognition (Biggs 1985) and epistemological sophistication (Bing and Redish 2012). Both of these constructs deal with students' awareness of themselves and how they learn. The *achieved* and *moratorium* identity statuses have a large amount of exploration as a core condition of the status, and the argument could be made that awareness of oneself and how one learns could be part of this exploration process. It makes sense then that the more exploration one has undertaken, the more likely that an individual has become more meta-cognitively and epistemologically sophisticated through this exploration.

The identity that is being focused on in this paper is subject-specific identity. We apply the identity status framework to examine students' physics identity. The crises in the context of this study are related to college major decisions or pursuance to particular subject-specific experiences such as an undergraduate research experience in physics. The identity status framework is relevant to the investigation of students' decision to study a subject in any context but is particularly relevant to investigations in the United States where the college major decision is an important life focusing decision (Galotti et al. 2006) that many students struggle with (McDaniels et al. 1994). The theory fits the United States college context, as the college major decision is open to both multiple approaches and multiple sources of information that can be investigated to help inform this decision. The career domain which is relatable to the major choice or subject choice is a core component of one's overall sense of identity (Kroger et al. 2010), and Veerle Germeijs and colleagues (2012) have argued that Marcia's model is applicable when dealing with specific identity relevant decisions in one domain (i.e. subject-specific identity).

**Communities of practice and identity statuses**

An area of convergence between identity statuses and communities of practice is the period over which these constructs can influence a student. When a student attends college she will both intentionally and unintentionally join several communities of practice ranging from classroom communities to the professional communities of practice of the subject they are pursuing (Irving and Sayre 2014). One's place within these communities of practice changes over time. When a person initially enters a community of practice, they enter as a newcomer who is gradually socialized into the community of practice through mutual engagement with and support of old timers (Barab et al. 2002). Through low-level but authentic practices, these peripheral participants are slowly inducted into the knowledge and skills of a particular practice. Over time, they develop more understanding, knowledge, and skills, becoming central participants and eventually mentoring their own peripheral participants (Lave and Wenger 1991). Students in the process of moving from being a peripheral participant to central participant are referred to as having a trajectory towards being a central member of the community.

Earlier it was discussed that, for the framework of identity status, the emphasis is on exploration and commitment in relation to deciding about one's identity at crisis points. The communities of practice framework fits the identity status framework as a community of practice facilitates both exploration and commitment. Being a member of multiple, overlapping communities of practice is true throughout one's life, but especially true for an undergraduate in college. The undergraduate period offers the opportunity to explore multiple different identities.

Each one of these communities of practice then provide opportunities to display one's commitment to an identity by engaging in the significant practices of that community or several communities that relate to an identity choice. Hence commitment is in evidence when students begin to become more central members of communities of practice that are related to their identity choice. James Côté and Charles Levine (2002) found that individuals in an achieved identity status were more likely to self-assess themselves as integrated into their respective professional community. Koen Luyckx et al. (2013) argue that individuals in achievement status also tend to find a validating community for their professional identity and do so at a faster rate when compared to individuals with diffused statuses. The relationship between identity status and community of practice is also evident in the many possible crisis points brought about by membership in communities of practice. Decisions such as joining a community in the first place or whether or not to take up more central practices in the community when that opportunity arises are potential crisis points for students.

**Affect and identity**

Identity and affect, when discussed in the context of learning and teaching, are inextricably linked (NRC 2009). This link is especially evident when discussing subject choice and the factors that influence this choice. The axis of exploration and commitment discussed in relation to identity statuses is influenced by the affective domains (Boe and Henriksen 2013). If we examine subject choice from the perspective of science and physics in particular, the affective domains have an important influence on both attitudes toward science and performance in science (Hazari et al. 2007). Steve Alsop and Mike Watts (2003) argue that particularly at moments of choice the affective and cognitive domains are greatly interrelated. The range of affective factors that can influence the choice to pursue physics or the choice to persist with physics is vast: encouragement from other people (Hazari et al. 2007); students' situational interest in physics; personal interest in physics; importance of physics; physics course anxiety; physics test anxiety; physics achievement motivation; student motivation in physics; self-efficacy in physics; self-concept in physics (Gungor et al. 2007); and attitude towards physics (Simpson et al. 1994). Of these affective factors, self-confidence and self-efficacy are often indicated as the most significant and powerful influence on persistence (Wigfield and Eccles 2002) with a subject and hence at points in time when the choice to persist manifests.

**Affect and physics learning environments**

Emotions are a central part of learning and teaching. For example, the affect that occurs from one's self evaluation of their ability in a subject can have a deep effect on their pursuance of that subject, but research into affect is minimal (Alsop and Watts 2003) and centers mostly on motivation. The ordinary physics student is motivated by their interest and enjoyment when it comes to physics (Carlone 2004). Affect, particularly interest and enjoyment, is visible in students' behavior in physics learning environments. Affective responses include verbal expressions of feeling, gesturing and body language, and tone. They are common in physics (Allen 2002), occurring during most learning situations in most learning environments. Affect can positively, neutrally, or negatively influence the progress of the learning process (Hassi 2009). With this range in influence, the affective domain can, therefore, significantly enhance, inhibit, or even prevent student learning in the sciences (Simpson et al. 1994).

Positive affect promotes students' willingness to take on difficult challenges and to think creatively when confronted with complications (Hunter et al. 2007). Students with heightened interest in physics are more likely to pose curiosity questions and use systematic approaches to seeking answers (Engle and Conant 2002). Interested students are also more likely to use effective learning strategies and reflect on feedback (Barron 2006). Other positive affective responses include joy in tasks, inspirational exclamations during "aha" moments, and increased camaraderie.

Affect frames how students relate to fellow students (Maulucci 2013). We can observe the resistance and uptake of classroom norms through students' interactions which have strong affective aspects. Affective states (mood) can also influence the information shared in laboratory environments (Eich et al. 2000). In collaborative learning environments, the combination of multiple interests and perspectives can lead to visible tension and conflict within a group or individual (Sins 2010). In contrast, low-conflict groups may promote positive affective responses. For example, in the free time between tasks, students may make spontaneous positive proclamations about their experience of learning physics.

**Research goals**

Our research goals in this study are:

- To ascertain the identity status of each student through their discursive and affective responses.

- To map the three students' pathways within physics as they gain various experiences and change identity statuses.

- To relate students' identity status with their trajectories towards becoming central members of the community of practicing physicists.

**Methods**

The research presented in this paper is part of an ongoing ethnographic research project on the identity development of undergraduate physics students.

As a methodology, ethnography originates in anthropology and is concerned with the sociocultural features of the environment, including how people interact and their discursive practices (Brown 2004). In educational settings, it is used to characterize relationships and events that occur in different educational settings (Gordon et al. 2001) and results in rich descriptions that make it possible to understand what is happening and why (Collins et al. 2004). Ethnography typically draws its data from a number of sources in order to get a complete picture of the events and relationships being characterized, but also in an attempt to overcome some of the weaknesses of subjectivity through triangulating multiple viewpoints (Yin 2011). Our data encompasses two different sources (observations and interviews) to triangulate multiple viewpoints on students' developmental paths towards or away from physics (Stevens et al. 2005).

Instructional contexts

The students in this study are enrolled in upper-division undergraduate physics courses at a large research institution in the middle of the United States. For the most part, their courses are typical of courses at that level across the continent, enrolling 5-30 students per course. While students are not organized into coherent cohorts, they do tend to follow similar sequences of courses over time, and cohort-like effects develop naturally.

Data Streams

The primary source of data for this paper comes from semi-structured ethnographic interviews with three students recruited from upper-level physics courses in electromagnetism, mechanics, modern laboratory, and advanced laboratory (AdLab) as part of an ongoing identity study. We developed a 45 minute semi-structured ethnographic interview that was structured in a past-present-future format (Stevens et al. 2008) in which students reflect on what they used to think, currently believe, and anticipate for the future. The semi-structured nature of the interview means that the interviewer has no set script but has several talking points which are drawn from literature on identity formation (Aschbacher et al. 2010), epistemological sophistication (Bing and Redish 2012), and metacognition (Sandi-Urena et al. 2011). We interviewed each student at three different points in their academic careers, following each student regardless of his or her course sequence. The timeline figures (Figure 2, Figure 5, and Figure 8) indicate the period in the students' academic careers that each interview took place. The interviews were videotaped and transcribed for analysis.

A secondary source for the ethnographic analysis comes from observations of students participating in AdLab. The AdLab classroom involved students doing structured experiments over a period of two weeks at a time before rotating to a new experiment. The experiments typically involved physics concepts that the students have been previously introduced to in prior classes, and tend to be classic experiments from the $20^{th}$ century, e.g. Millikan's measurement of the charge-to-mass ratio of the electron, or the half-life of muons from cosmic ray showers. The experiments also involve setting up and using equipment that is more complicated than the students have previously encountered.

Within the lab classroom, groups of three students worked together at each of approximately six experimental set-up areas. Each group worked on a different experiment at a time. We video-recorded the groups twice per week for three hour class sessions for a full semester (Irving and Sayre 2014). No one can observe identity, but what can be observed is the behaviors that should result if an identity is fully formed, partially formed, or has not been formed at all. The purpose of the physics laboratory is considered to be an enhancement of student mastery or conceptual understanding, development of scientific reasoning or practical skills, illustration of the nature and complexity of scientific work to students, and cultivation of student interest in the subject and the development of group work skills (National Research Council 2005). Due to the multifaceted nature of the learning goals of the physics laboratory (including the AdLab environment) and the collaborative nature of the environment, it is a rich source of observable affective behavior. Affective behavior is one way of observing where in the formation process of a subject-specific identity a student resides.

**Analysis**

Initially, the analysis process started at a macro-level where we examined all of the transcripts of all of the interviews together to identify segments of the interviews that shared thematic content.

The focus was not solely on the transcript and instead involved working concurrently with the transcripts and videos of the interviews as the transcripts often lacked the depth of completeness that the videos contained. We took a traditional case study approach which allows the examination of how the students and their responses change over time while also allowing for comparative analysis between the individual students (Hammer 1994). The initial macro-level ethnographic analysis allowed for the identification of thematic content that could then be mapped over time (Stevens et al. 2008) and between subjects.

One theme that emerged from our data was how the students either did or did not self-identify as physicists. The past-present-future format of the interviews allowed us to explore how this theme did or did not change for each student over time. Another theme that emerged was how their professional aspirations changed in response to their experiences. The use of the past-present-future format of the interviews enabled us to map how the perceptions of professional aspirations changed over time. (Other themes emerged; which are not related to the focus of this paper and will not be discussed here.) We selected the chosen themes for further study and analysis to help us explore what identity status the students were in and how this status manifested in the students' interactions with strong affective aspects in AdLab. The next step of the process was to use micro-ethnographic analysis to begin exploring each theme for each student in greater detail. Statements about one theme were analyzed and compared to statements they made related to the same theme during another part of the interview or in an alternative interview. The researchers then sought an understanding as to how the theme being analyzed fit into the context of the student's relationship with their identity as a physicist and hence their identity status. Once this process was complete, we analyzed the theme as a whole with the other students' interviews and we then grouped statements about similar experiences or perceptions together in order to complete a comparative analysis between the students. With both thematic analysis and comparative analysis completed, we constructed case studies of each of the three students (Cresswell 1994) to describe each individual's identity status and their evolving relationship with their physics identity.

We conducted the analysis for the observational data in a similar manner. The focus was on identifying themes of an affective nature in the student's interactions with their group members and other students within the classroom. When we identified interactions with strong affective aspects, we would examine the context and content of the interaction. We considered the pre- and post-context of the interaction, their speech and body language was used to interpret how the participants framed the interaction. In relation to body language, the students' posture, gaze, and gestures were analyzed in the context of the laboratory environment for indications of positive or negative affect whilst completing an experiment. Posture can provide evidence of a student's availability for entry into or out of an encounter (Goodwin and Goodwin 2001) and gaze can mark states of engagement and disengagement (Goodwin 1981) while gestures can provide an insight into a student's affective state (Fast 1988). How they framed the interaction refers to the resources the students bring to bear for a particular interaction (Irving et al. 2013). We were interested in the affective resources that are activated and the resulting affective responses that occur in the student discourse (NRC 2009). This analysis of themed interactions with strong affective aspects in conjunction with the ethnographic approach allowed for correlation to see how interactions related to a student's identity status. The final step of the analysis was to triangulate the themes identified in the ethnographic analysis of the observation data to those identified in the interview data.

Initially, the interview and observational data were coded concurrently and independently by two researchers to ensure the reliability of the themes identified. This process also acted as a validation process of the interpretations being made from the identified themes. After initial coding the researchers met to compare the themes they had identified and their interpretation of the meaning of the themes identified. Although there were differences in interpretation of themes, there were very few instances of themes not being identified by both researchers. For differences in interpretation a negotiation process occurred until agreement was met. This negotiation process often resulted in the researchers reviewing the interviews or observational video data again in order to provide evidence of their interpretation. After the themes and interpretations had been co-identified and negotiated a single researcher wrote up the analysis of the case studies with the other researcher "checking-in" on the analysis by reviewing sections of analysis. The final reliability and validity check was presenting the research to the physics education group at Kansas State University in order to conduct a peer review (Lincoln and Guba 1985). The peer review focused on the group challenging the researchers' assumptions and asking the researchers questions about method and interpretation. After this review minor changes were made to each case. In the following section, the case studies are presented. For each student, the interview and observational data is presented in chronological order, as the time of each event is relevant when studying a construct such as identity status. We present several extracts and episodes, which give substance and support to the themes identified. In each case, they are included in the relevant parts of each case study. Words that relate to affect in an extract will be underlined in order to highlight them, as they will often be the focus of discussion in the following paragraph. In the following cases pseudonyms have been used for both the interviewees and any individuals that have been referenced by the interviewees. A typographical note concerning transcript in this paper: we use a period (.) to indicate a short pause, ellipses (…) to indicate a longer pause, and underlining to indicate emphasis. Items in square brackets ([]) are inserted into the transcript to improve readability.

**Case 1: Sally**

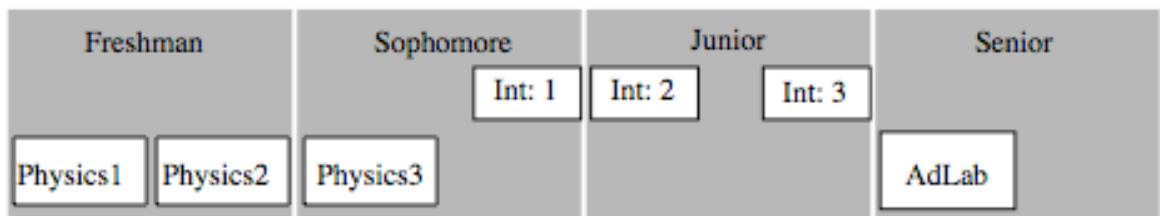

**Figure 3**: Data collection timeline for Sally. Physics 1 and 2 refer to Figure 2: introductory physics 1 and 2, which cover mechanics and electricity and magnetism. Int 1, 2, and 3 refer to the different interviews conducted with the student. AdLab refers to the advanced laboratory class.

Sally was first interviewed at the end of her sophomore year and was 20 at the time. Her first interactions with physics that she remembers are in middle school learning about energy. She took physics in high school but had a "horrible" teacher whom she described as not understanding the physics being taught and as a result recalls it being a mass of equations with insufficient background explanations. In high school Sally also took chemistry and after a positive experience with chemistry decided to double major in chemistry and secondary education. As part of her chemistry degree, she had to take two introductory calculus-based physics classes for physical scientists and engineers, Physics1 and Physics2. Sally had a much better experience with physics than the one in high school. She explains that the equations no longer were coming from nowhere, and the use of demonstrations allowed her to understand and connect the material to a much greater extent in real life. The condition of one of her scholarships is that Sally has to teach high school for four years after finishing her degrees. After Physics1 and Physics2, Sally decided to take the class Physics3 which was not a requirement, after which she decided to declare a physics minor. At the time of her first interview, she is taking mechanics and has just joined a physics research group specializing in Atomic, Molecular, and Optical physics (AMO).

1st interview

At the time of the first interview, when comparing physics and chemistry, Sally is classified as being more central in her trajectory to becoming a practicing chemist. Sally is also more committed to a subject-specific identity in chemistry. In the context of an academic career in the United States, the opportunity for exploration of another subject can become limited but not impossible as students begin to take upper-division courses. Sally had already decided to explore a subject-specific identity in physics by choosing to take Physics3 after taking the compulsory Physics1 and Physics2 classes. There is further evidence of this exploration as she has decided to take the Classical Mechanics course at the end of her sophomore year with the intention that it enables her to take quantum mechanics at a later point. Choosing to take both of these classes furthers her trajectory toward being a more central member of the undergraduate physics community of practice as it allows her to engage in the legitimate peripheral practices of attending more advanced classes, being examined on more advanced topics, and interacting with other physics majors.

*S1.1 Commitment to physics.*

Int:   So why did you choose to do that (take classical mechanics)?

Sally:   I want to take quantum mechanics and it's a pre req, mostly the only reason I took it.

At this point in her academic career, Sally has already made strong commitments to both chemistry and secondary education and to a future career of teaching high school. This would signify that she concurrently has three trajectories towards central membership in three different communities of practice that relate to a professional identity. She also has concurrent trajectories towards central membership in three different communities of practice that relate to a subject-specific identity. The decision to pursue the study of quantum mechanics that involves a prerequisite class and is a year away displays evidence of Sally deciding to explore physics as possibly another identity. It also highlights her interest in a particular area of the study of physics as she seems disinterested (at the time of making the decision to take mechanics her feelings towards mechanics are of disinterest, which subsequently change to be more positive) in mechanics as a subject and instead it is a means to an end. This level of reflection would seem to indicate an informational identity processing style, which is typically associated with a moratorium, or identity achieved identity status. This pursuit of quantum mechanics is also a result of the positive experiences Sally had in trying out her physics identity in Physics3.

*S1.2 Future plans.*

Int: Given that you have such a love for it, do you see a career path in it?

Sally: I'm hoping, <u>tentatively</u> hoping, right now I have to teach high school for a while because it is one of my scholarships, but my goal after that is to come back and get a doctorate in either physical chemistry or chemical physics in there, something with quantum mechanics. I'd like to do research at a university.

Sally is not making a commitment at this point - it is just a suggestion of a commitment Sally would like to pursue when there are no longer any restrictions on her choices. There is a level of hedging in her hope as indicated by "tentatively". Aspiring to a doctorate in physical chemistry or chemical physics indicates a trajectory towards a professional identity in the community of practicing physicists or chemists. In summation, Sally has already made commitments to two subject-specific identities: chemistry and secondary education. At this point in her academic career, Sally self identifies as a chemist, and we would argue that, at the time of the first interview, Sally's primary subject-specific identity is a chemist. However, she has now implemented an exploration of physics.

Sally's discourse during the first interview is full of affective responses especially when talking about physics.

*S1.3 Positive affect.*

Int: What is it about quantum mechanics that interests you?

Sally: It's just so <u>beautiful</u>. I just <u>love</u> the way it describes the world at a basic level. <u>As a chemist,</u> I <u>love</u> the atomic scale… that is where I can see best…and to be able to see it explained so clearly… Stuff that like you just can't learn in a Chem1 class cause you just aren't at the level. To see what is actually going on, um, was just <u>beautiful to me</u>, I just kind of <u>fell in love with it.</u>

The previous extract is very interesting as Sally is not only talking about her love of a particular physics subject area, but she also makes a comparison between chemistry and physics. She describes how chemistry does not allow for a complete explanation of the material for which she has such an affinity. Sally makes a similar comparison when she compares physics to chemistry research.

*S1.4 Comparing physics to chemistry.*

Sally: Like you spend so much time designing an experiment and trying to get everything right and trying to see this phenomenon that you're expecting… and you spend so much time working at getting everything in the experiment to work and you finally get it to work and you <u>get really excited</u>. But in chemistry you just mix solutions together over and over again and you get stuff, I don't know, it's just <u>not as exciting</u> to me.

Sally is exploring both her subject-specific identity in physics and her possible professional identity in physics through engagement in legitimate peripheral practices of both communities of practice through the activities in her physics classes and through working in her physics research group. When describing her experiences in both, she communicates a positive attitude and substantial personal interest in physics.

At this point in Sally's academic career she has already made commitments to two subject-specific identities: chemistry and secondary education. She is in the process of exploring physics. To date, her exploration has been positive and is encouraging her to make a bigger commitment to the subject. Sally's physics subject-specific identity is not currently her primary subject-specific identity and is currently in a *moratorium* identity status. The same is true for her professional physics identity. Sally's commitment to physics is limited, and she is still in the process of exploring it.

2nd interview

By the time of the second interview, Sally is now in her junior year and has worked in her physics research group for the duration of the summer. She has found it to be a very rewarding experience. Sally still feels a sense of uncomfortableness as she still does not have a complete understanding of the area her research group conducts their research in and of the research that was carried out during the summer. Sally does feel, however, as if she is closing the gap at least with the graduate student members of the research group.

   *S2.1 Joining a community of practicing physicists.*

Int:   Do you feel comfortable talking to the other members of the research group more now than you previously did?

Sally: More now, yes definitely… Not super comfortable still, especially when they ask me questions about what I am doing…definitely more comfortable than I did before.

Sally's continued membership and engagement in the activities of the research group is a further exploration of both her professional physics identity and her subject-specific identity. The new development is that she has progressed along her trajectory to be a more core member of the community of practicing physicists by engaging in peripheral legitimate participation of the community of practicing physicists. She is now engaging in central practices of the physics community. By engaging in these activities she has also become a more central member of the undergraduate physics community of practice. She is also becoming more accepted by her colleagues whom she considers physicists. Sally also explored one of her other subject-specific identities during the summer which resulted in reflection and reinforcement to her commitment to physics. Again Sally's reflections on her experiences, commitments, and the information she has available to her relates to an informational processing style which in turn is related to a moratorium identity status.

*S2.2 Exploration of identities.*

Sally: Simultaneously while I was working the research lab this summer, I was also doing education classes. I was getting to teach students small lessons. It made me realize as much as I loved teaching… teaching middle school or high school, I wouldn't be satisfied. Like, I would get <u>bored</u>, you know, teaching that for the rest of my life. Even though stuff changes I realized I needed more of a challenge.

Despite the firming of her commitment to the subject of physics, Sally is still not ready to self-identify as a physicist. During the summer she has also engaged in the legitimate peripheral practices of a chemistry teacher and has become dissatisfied with the trajectory towards central membership of this community practice and therefore it as a professional identity.

*S2.3 Perceptions of what it means to be a physicist.*

Sally: (laughs) I think that's kind of presumptuous of me to say that, because I don't even have a degree. I'm not even a physics major. I would love to be able to consider myself a physicist. If I was to declare a physics major, then I would feel more comfortable… because then I am working towards a dedication to physics, and it's obvious I have a dedication to physics.

Sally herself admits that she has not fully committed to either her subject-specific physics identity or her professional physics identity and, therefore, continues to occupy a moratorium physics identity status for both identities. In this interview, Sally continues to communicate positive affective responses when talking about physics and these become interwoven with statements about identity.

*S2.4 Positive affect.*

Int: In that case then, when you're at a social engagement and someone asks you what you do, what do you tell them?

Sally: I tell them I'm a chemist, and it makes me <u>really sad</u> because I want to say I'm a physicist.

Int: Is there any group of people that if that question did come up you would feel comfortable calling yourself a physicist?

Sally: I think… I think my closest friends would understand because they have seen me <u>struggle</u>... wanting <u>so badly</u> to declare a physics major, but just not having the time in my schedule. And they see how much <u>I love it</u>, and so they know how dedicated I am to it.

In the above extract, a preference is again displayed towards physics over chemistry. The final part of the extract is particularly interesting as previously Sally has talked about dedication, and that dedication to physics needs to be perceived and validated from some external group. In extract S2.4, Sally indicates that she does show physics the requisite dedication but that she cannot declare a commitment to the subject due to the time constraints of her busy schedule. In extract S2.1 Sally displays both self-efficacy in the assessment of how well she is doing in her interactions with her group members and anxiety about how well she fits into her community. This anxiety does not dampen her interest or love for the subject.

3rd interview

Between the second and third interviews Sally officially adds physics as a major and so has committed to her subject-specific physics identity. In this interview, Sally is near the end of her junior year and reveals that she has reflected on her experiences to date, including the interviews conducted. A result of this reflection is that she has decided to take physics as a major.

*S3.1 Crisis point.*

Sally: It was combination of a lot of factors. Part of it was actually this interview made me think, like I really enjoy physics, but I'm not willing to describe myself as a physicist and why not? I just spent some time thinking about that and realizing the place that physics had in my life that I hadn't fully accepted yet. Thinking about it made me realize how much I really enjoy it and actually made me consider doing physics for... you know… the rest of my life.

In relation to her research group Sally has also made progress:

*S3.2 Joining a community of practicing physicists.*

Sally: I think they (other researchers) view me as more of a physics student than before. Before it was just… she's a chemistry student who is dabbling in physics. I view them as equal in my mind now.

From extract S3.1, it would appear that physics has not only become Sally's primary subject-specific identity but also her primary professional identity: physics research "actually made me consider doing physics for… you know… the rest of my life". It is also clear from extract S3.2 that Sally is no longer feeling the anxiety that was evident previously in her interactions with the other members of the research group. Having made a commitment to the subject, Sally feels more accepted by her physics research community. Becoming accepted by her research group and the sense of cohesion she feels with the group provides a shift towards more central membership in the community of practicing physicists. This also indicates a point in which the legitimate peripheral practices of the community of practicing physicists and the undergraduate physics community of practice begins to diverge. Becoming a central member of a physics research group and engaging in the practices that this membership entails is not a requisite to become a central member of the undergraduate physics community of practice. Sally is also still exhibiting the same positive affect towards physics that has consistently emphasized her enjoyment of the subject. By the final interview, after much self-examination and reflection, this change in commitment indicates that Sally has changed her identity status to identity achieved in relation to both her subject-specific identity and her professional identity.

Observations

After the interviews, Sally joined the AdLab classroom in which we took observational data at the beginning of her senior year. In this environment, Sally continues to display her positive affect for physics. However, now these opinions are spontaneous and unsolicited.

In the following extract, Sally's group is coming to the end of working on the torsional oscillator experiment and is in the middle of discussing the experiment that they would like to do next. A torsional oscillator is something which twists and untwists rhythmically, oscillating in a circular fashion. Modifications on the experiment include adjusting the length and mass of the oscillator to change its frequency, and adding fins in order to investigate the effects of air resistance on its motion. Compared to other experiments in the lab, this one is technologically simple: the controls are straightforward, there are no complicated electronics to set up, and the signals to measure are visible to the naked eye. However, these experimental affordances also mean that it is one of the least "flashy" experiments: the theory it tests is classical mechanics, and there's little experimental cachet to making it work.

In this extract, Sally has an affective response to how theory and experiment work together, including in the torsional oscillator.

*S0.1 Positive affect.*

Sally: <u>It's cool</u>. No matter which lab you do, even torsion oscillator, you get to see stuff in action you read about, you know. Like you get to see theory working and <u>that's just cool. Who doesn't like that? And if you don't like that, there is no reason for you to be in this class.</u>

Sally is once again displaying her positive attitude to physics and her personal interest in the subject. More importantly she is now displaying evidence of having become a more central member of the community of practicing physicists and the undergraduate physics community of practice. From her previous comments, which indicated that Sally hid her commitment to physics in the past, it is hard to believe that she would make a comment like "if you don't like that, there is no reason for you to be in this class". This comment would indicate that Sally is speaking of some of the requirements for membership in the community of practicing physicists and the undergraduate physics community of practice in which she is now a member. Through her research experience and declaring physics as a major, she is now acting as a central member.

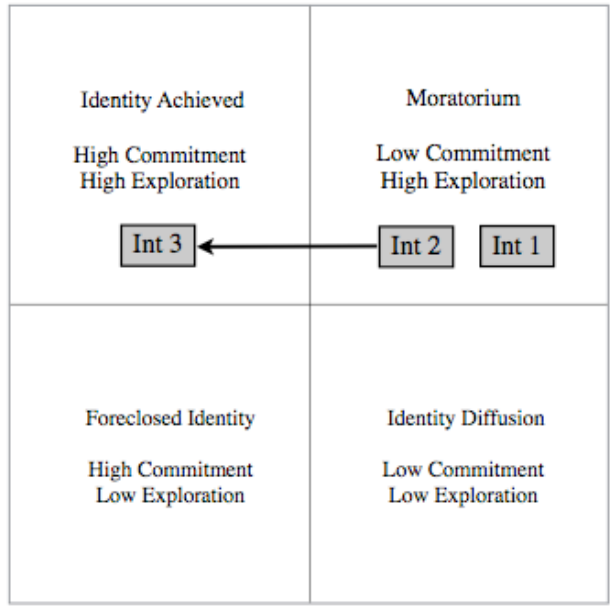

**Figure 4**: Sally's changing identity status for her subject-specific identity with each interview.

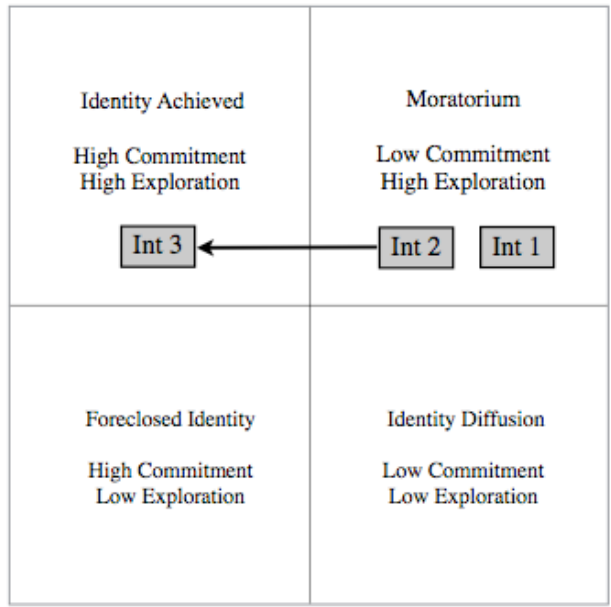

**Figure 5**: Sally's changing identity status for her professional identity with each interview.

Figure 4 indicates the identity status that Sally occupied from interview 1 to interview 3 in relation to her subject-specific identity and her professional identity. As can be seen Sally remained in *moratorium* for the first two interviews as she explored her identity as a physicist. After exploring her identity as a physicist through legitimate peripheral practice in both communities of practice Sally then decided to declare a major in physics and as a result her subject-specific identity and her professional identity switch to an *identity achieved status*. Sally displays evidence through her constant reflection on her preference for physics, of possessing an informational identity processing style which correlates with her moratorium and identity achieved statuses.

**Case 2: Larry**

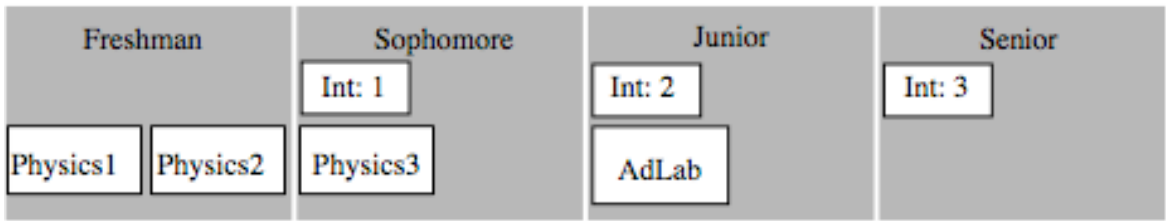

**Figure 6**: Data collection timeline for Larry. Physics 1 and 2 refer to introductory physics 1 and 2 which cover mechanics and electricity and magnetism. Int 1, 2 and 3 refer to the different interviews conducted with the student. AdLab refers to the advanced laboratory class.

In contrast to Sally, Larry is majoring in physics only. At the time of the first interview, Larry is 19 and a sophomore. Larry became interested in physics early in middle school when his teacher used to show videos of Bill Nye, "the science guy" the presenter of a popular long-running child centric science show in America. Since then, Larry became fascinated with Neil deGrasse Tyson and has read many of his books and watched many of his programs. Neil deGrasse Tyson is a physicist who has become popular through his many appearances on television shows and science documentary specials in America and abroad, particularly after Cosmos. In Larry's discussion on who may have influenced his decision to pursue physics, he points to Neil deGrasse Tyson's influence. In particular he points to Tyson's pursuit of presenting complicated physics concepts to the public. Larry indicates that Tyson "loves to help everyone understand" on several occasions throughout the interview and this is an attribute he believes he shares. Larry took physics in his junior year of high school and advanced placement physics in his senior year of high school. Although he found it interesting, it did not live up to his expectations as it did not touch upon the modern applications of the subject and did not cover anything in any great depth. He references a time that he had to research an English paper on the Large Hadron Collider (LHC) in high school as a major moment in his history with physics because each new idea he uncovered during his research "blew his mind". During the research for this paper, he often found he would not understand the particulars of many aspects of the LHC research and so started reading scientific papers in order to try and obtain an understanding.

Similarly to Sally, Larry talks about physics being much fun and also indicates a passion for the subject. He also emphasizes the difference between high school physics and college physics, indicating that college physics is much more interesting and covers topics much more comprehensively. He also likes the emphasis on developing problem-solving skills. He calls himself a "visualizer," meaning that he understands concepts and problems much better when he can see them, and he feels his physics classes so far have facilitated this preference. Larry participated in three interviews at the beginning of each of his sophomore, junior, and senior years.

Interview 1

In the first interview the discussion focused on several topics but the most pertinent topics to this study are highlighted with extracts in the paragraphs that follow. At the time of the first interview Larry has a *foreclosed* subject-specific physics identity and a *diffusion* status in relation to his professional identity. He has made a commitment to obtaining a degree in physics, but is unsure as to what to do with this degree. When compared to Sally, Larry does not declare an overriding interest in a particular physics topic as Sally does with quantum mechanics nor has he indicated that he has explored the various areas of physics that are open for him to investigate. He has made no other physics-related commitments other than making a commitment to pursuing a degree in the subject. His most recent crisis point of choosing to pursue physics in college has resulted in the beginnings of a physics subject-specific identity being formed.

  *L1.1 - Future plans.*

Int:   What are your plans career-wise?

Larry: I don't have a definite plan. Recently I have been talking about how <u>I love</u> talking to people about physics. That might be a good indication that I should go into teaching…I don't really know what I want to do. I do know I want to get my degree in physics. That is something I'm pretty sure at this point. I don't know… I might decide, as the semester goes on… I might decide this professor is doing really <u>cool research</u> - I would like to do something like that! Or, you know, find out about a field of study in particular that I didn't know much about or something like that. I'm trying to keep my options open. At <u>some point I'm going to have to decide</u> exactly what I want to do.

Many sophomores — like Larry — don't have concrete career plans yet; however, Larry's statement also indicates a lack of exploration around future plans in physics. Larry's lack of exploration is also evident by his lack of desire to pursue a particular field or his opinion that a research experience can be postponed until it can fit around his physics and band schedule (he plays in the college marching band). He also refers to the fact that he could pursue a math minor — "it would be easy for me to get a math minor after all my required courses and so I might add that on but I'm not sure though." Yet again there is no assuredness or commitment to this as something worth exploring. What he is sure of is that he is committed to physics, which indicates his subject-specific physics identity. The underlined sections in extract L1.1 also indicate positive affect towards the subject of physics that is heightened in his indication of a love of talking about the subject. The next two extracts continue this theme of positive affect towards the subject and talking about the subject of physics.

*L1.2 - Positive affect.*

Int: How are you finding it (physics) so far?

Larry: <u>It's really fun</u>. So far….I like both of them (Physics1 and Physics2.) I have learnt so much more than I thought I would….but yeah physics I am <u>definitely happy</u> with it so far.

At this point in Larry's physics career he is very enthusiastic in his choice to pursue the subject as his major and his experiences so far have been very positive. Larry displays similar positive feelings as Sally did towards physics throughout the first interview.

*L1.3 - Positive affect continued.*

Larry: Different people have different levels of understanding, and so I try and do what the professors do, which is establish how much they know and do my best to explain from there. But as far as that's concerned, I'll talk about it with anyone pretty much as much as I can. <u>I love talking about it (physics)</u>.

The above extract is of particular interest as it highlights Larry's affinity for the subject and also an affinity for his teaching of the subject. He displays evidence for his appropriation of his professor's approach to teaching. The appropriation of his professor's approach to teaching is an example of Larry being guided by his professors in a legitimate peripheral practice of the community of practicing physicists and hence giving some acceleration towards central membership. However, it would not be inappropriate for this practice to also be a legitimate peripheral practice of the undergraduate physics community of practice and so he has also been given some acceleration towards central membership of this community. From a professional identity perspective Larry is also exploring the idea of becoming a teacher, possibly unbeknownst to himself. Despite this subliminal exploration, Larry would still be classified as having a *foreclosed* identity status in regards his subject-specific identity and a *diffusion* identity status in regards his professional identity. This is evidenced by his combination of a small amount of exploration matched with a commitment to pursuing a degree in physics. Evidence of appropriation indicates that Larry is getting a taste for the practices of the community of practicing physicists, but it is interesting that the legitimate peripheral practices he is appropriating are focused on teaching approaches. This ties into Larry's perceptions of being a physicist, as he believes that being a physicist is a mindset and that what he has learnt is a mindset, including the way of breaking down a concept to teach it.

*L1.4 Perceptions of what it means to be a physicist..*

Int: Are you a physicist then, given that?

Larry: I would like to think so. I might be a little bias...I like to think I am.

Int: How long have you been a physicist?

Larry: Definitely since enrolling in EP1 (Engineering Physics 1). Being in that class… I think, if you want to succeed you have to be in that mindset.

Larry considers that being a physicist is a matter of mindset as opposed to Sally's interpretation that it is a matter of commitment. Due to this interpretation of what it means to be a physicist, Larry perceives himself as being a physicist already. Larry's perception that he is already a physicist is consistent throughout all three interviews and this belief could make Larry feel as though he does not have to explore or gain other physics experiences beyond those provided by his studies. This is indicated in the next extract in which he remarks that research is not necessarily an important part of being a physicist.

*L1.5 Research.*

Int: What makes a physicist to you?

Larry: Well first off, you have to be involved in physics somehow. I don't think you necessarily have to be, like, doing research.

This would indicate that in relation to the various physics-orientated communities of practice, Larry is more focused on engaging in the legitimate peripheral activities of the community of physics undergraduates. Larry does not feel that the legitimate peripheral activities that are orientated more towards the community of practicing physicists are as relevant to him besides those that are orientated towards teaching. A student can engage in the legitimate practices of the undergraduate physics community of practice without ever having to engage in the more central practices of community of practicing physicists. Larry's perception of what it means to be a physicist results in him ignoring the guidance of central members of the community of practicing physicists and the need to participate in a legitimate way beyond the legitimate practices that both the community of practicing physicists and the undergraduate physics community of practice share. Larry has and is still attempting to appropriate the proper teaching-related attitudes of physicists associated with teaching but is not engaging in the significant practices that Sally did, like doing research or seeking acceptance from a community of more physicist-like practitioners. He rejects the idea of research as being an important legitimate peripheral practice of the community of practicing physicists.

Interview 2

The second interview takes place approximately six months after the first interview, and it is near the start of his junior semester in which Larry is taking the AdLab class. Larry's identity status seems to have remained unchanged since the last interview, as he seems to still be in a *foreclosed* subject-specific physics identity and a *diffusion* professional identity. This is evidenced by him still being undecided about his future career and he continues to express a minimal preference for teaching but has not yet made any active attempts to pursue this path or make a commitment to it. At this point, he has a minimal commitment to pursuing a teaching career and a strong commitment to gaining his physics degree and is currently contemplating exploring his option of becoming a researcher. Contemplation of exploring research is pretty far from actually exploring research.

*L2.1 Research.*

Larry: I'm not as <u>interested</u> in research… like, the advanced lab class is going better than I thought it would. I wasn't anticipating enjoying it at first. I think research has the potential to be <u>very frustrating</u>, I just feel research being non-idealized could just be <u>really stressful</u>, a sort of <u>aggravating process</u>, but I'm considering trying to work with a professor sometime in the future to get more experience and I might change my mind after I try working in a legitimate lab setting, especially with someone who really knows what they are doing.

Larry's experiences in the AdLab environment seem to amount to a pseudo-exploration of what research might entail. The AdLab environment models some of the legitimate peripheral practices of a professional physicist with its emphasis on two-week projects that allow investigations with advanced equipment and writing reports to demonstrate experimental findings. Larry's demonstrated hesitance and negative affect towards research and its relationship to the AdLab environment has resulted in him not being motivated towards engaging in other legitimate peripheral practices of the community of practicing physicists. In this interview, Larry again talks about getting a research experience like he has in the previous interview, but has not gotten around to achieving such an experience yet. Despite his experiences in AdLab being described as positive, he still has an obvious negative disposition towards research as a career option. He uses affective responses like "very frustrating," "really stressful," and "aggravating process" to describe research. These statements are particularly germane, as Larry believes that the AdLab learning environment is a more authentic experience of what "real science" might be like. He is using an unauthentic experience as a measure of his possible appetite for pursuing physics as a future endeavor.

*L2.2 Perceptions of what it means to be a physicist.*

Larry: Phil and I, in the Zeeman lab, we spent a long time just talking about… we'd take a measurement and we'd look at it and we'd say,' wait that doesn't look right, what might not be right about that.' It felt like real science… you know, it doesn't feel, like watered down, like some of the introductory physics classes do.

In fact, it is when he reframes the laboratory as a problem-solving endeavor that fits his description of physicists having a mindset and approach to problem-solving that he does talk about it in a positive affective manner.

*L2.3 Problem-solving as physics.*

Larry: It felt really cool to be faced with this difficult problem. Since you've got three hours, you have a long time to talk about or just figure out where you're going wrong or what calculation you may have or something like that… and it's really cool to sit down and work out a problem like that and figure out… like here we figure out it was the splitting of one half instead of one third or something like that… and so opportunities like that make this class pretty cool.

There is an interesting combination of perceptions occurring between these two extracts. On the one hand Larry equates his experiences in the laboratory as being the authentic activities of a scientist. There is an element of truth to this perception in that he is engaging in some of the legitimate peripheral practices of a practicing physicist. But these are not the same practices of the more central practices involved in carrying out research. Larry also reframes the legitimate peripheral practices that he engages in, in the laboratory, so that they are more aligned with the legitimate peripheral practices of the undergraduate physics community of practice rather than the community of practicing physicists. Despite his perception that he has had experiences that felt like "real science," Larry essentially feels he is in the same place regarding his subject-specific identity and his professional identity as he was six months ago. He indicates being closer to doing work in physics in the real world but his thoughts on what it means to be a physicist have not changed despite his description of his activities in AdLab as being more authentic.

*L2.4 Perceptions of what it means to be a physicist.*

Int: In any way do you think you are more of a physicist after taking the classes you have taken this semester?

Larry: That's a really hard question to answer, 'cause I'm getting…I'm certainly getting closer to being able to do work in physics in the real world, but I effectively have the same mindset and same outlook as I did last year. So I'm leaning towards no I guess.

Larry's decision making, lack of exploration, and commitments are resulting in an undefined career path or future relationship with physics. He does seem to be influenced by his positive emotions towards teaching physics. Conversely, Sally's decision-making, exploration, and commitments are being influenced by her affect towards one particular branch of physics and her experiences as a researcher and teacher. It is these experiences and emotions that allowed her to reflect and construct an appropriate pathway through physics for her intended career path. When asked what it was about teaching that appealed to him, Larry responded in the following manner.

*L2.5 Positive affect.*

Larry: Working with people, especially in engineering physics… I did homework with a group of other students a lot… and if I understood a problem, I always loved trying to help someone else understand it, and I'm not sure why, but I know that I did. I just really like the subject and I like to help other people to appreciate it and understood it as well.

Larry will eventually come to a crisis point where he will have to make a decision about his physics future but for now he is happy to let his emotions towards helping people understand physics and his class schedule guide him.

Observations

The analysis from the observational data supports a lot of Larry's statements about the AdLab environment and his negative and positive feelings about it. From a negative perspective, Larry indicates many times that the experimental process can be frustrating.

In the following extract the instructor has come over to help Larry's group set-up the experiment. The group has been working on setting it up for over an hour now and have encountered problems on several equations. As the instructor joins the group she indicates that she was going to help them earlier but got distracted by the other groups in the room.

*L0.1 Negative affect.*

Instructor:   I was going to do it at the beginning of class, but we all got so excited (talking to the entire room), so I didn't.

Larry:   I was pretty excited at the beginning. <u>This is getting frustrating now.</u>

At another time during the same experiment they now believe they have the experiment set-up correctly and are trying to take measurements but are finding it difficult to change the current being supplied to one of the instruments in even intervals as small changes on the dial are resulting in large changes on the read out.

*L0.2 Negative affect.*

Larry:  Labs can be so <u>frustrating.</u>

Bob:   Yeah...imagine being the first person to do this and they're just making it up as they go.

Matt:   You better have a lot of data.

References to frustration and difficulty are one part of Larry's typical affective responses to the AdLab environment. The other parts are positive affective responses as demonstrated in the following description of an episode that took place. Larry and his group are working on the derivation of an equation that is fundamental to the completion of one of the experiments his group has undertaken. Larry spends upwards of 40 minutes working on the derivation getting very excited at various stages until he has an "eureka" moment and throws his arms up in the air in excitement.

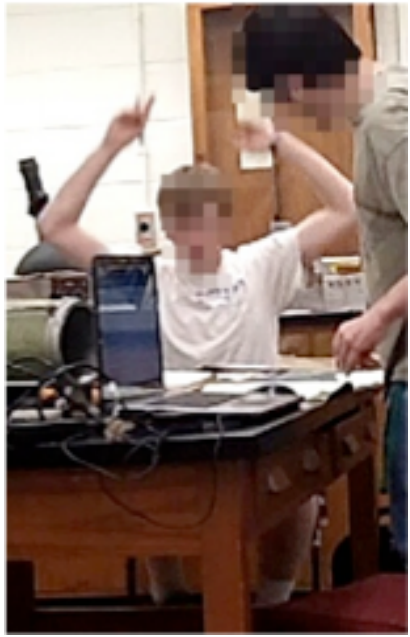

**Figure 7**: Larry's "eureka" moment. Larry is in the center; his lab partner, Matt, is to the side, looking at the equipment.

Positive affective behavior was common for Larry when working on problems in the laboratory that centered on math and took a large amount of time to solve. Overall the affective behavior reinforces the fact that Larry had both positive and negative responses to different elements of the laboratory. This again highlights that Larry seemed happier when the legitimate peripheral practices he engaged in were more orientated towards the undergraduate physics community of practice and not towards the community of practicing physicists. To reiterate problem-solving is definitely a legitimate peripheral practice of both a professional physics identity and a subject-specific physics identity, but it is a much more central practice of the undergraduate physics community.

Interview 3

Larry's third interview takes place during his second semester in his junior year. Larry is still happy with his choice to pursue physics as he states during the interview and he is still displaying positive affect towards the aspects of physics that he enjoys.

*L3.1 Positive affect.*

Larry: Yeah, like solving a difficult problem is always the best, 'cause you know… it feels nice to feel like you understand concepts, but when you can prove by solving a problem that you do get it, then you know you're making progress. I don't really understand why other people don't like physics. I gotta say, you know when somebody says 'you know, I

> hate that physics class,' I'm like 'Why?' But I'm not always successful at convincing people.

The above extract not only encompasses his continued positive affect towards problem-solving in physics, but also indicates that he does not understand when other students do not share in his positive attitude towards physics. As in previous interviews, problem-solving is the aspect of physics that he still enjoys the most, and he still ascertains that research or research experience is not necessary to be a physicist.

*L3.2 Research.*

Int: So research is not important then?

Larry: Yeah, I think so. It doesn't hurt your case if you're doing research, but I feel like, um, sort of like, as long as you are synthesizing knowledge, um then that's like the main criteria (to being a physicist). I'm not that interested in it (research). I don't feel like it's something I need to… that I really want to do. Then again I have never actually had a research position, so you know it might be once I do research I might go, this is great, why didn't I try this before?

This statement echoes Larry's previous interviews where he maintains the opinion that research is not that important and admits that it is not something that he necessarily wants to pursue. He also admits, like on previous occasions, that he is not basing these opinions on any real understanding of what an authentic research experience entails. He reiterates for the third interview in a row that he should gain such an experience in order to accurately understand whether it is something he might be interested in pursuing in his future.

There are some moderate differences in Larry since the previous interviews indicate that he might be on the verge of changing his professional identity status to a *moratorium* status. This transition is a change he has implicitly contemplated before. In Interview 1, Larry indicated a need to explore, when he talks about the need to try research or obtain more information about pursuing teaching as a career. Since the previous interview, Larry has actively sought to become a bit more informed about his career ambitions than he has in the past.

*L3.3 Exploration.*

Larry: Yeah, um, I actually talked with, I don't remember her name, but some sort of faculty member, because I think what I would like to do is get a teaching position after this, but I know I wanna leave college with a physics degree.

Larry is as certain as ever about his subject-specific identity but now has explored to an extent how to achieve his goal of becoming a physics teacher. The community of practice of physics teachers is a different professional identity than central membership of the community of practicing physicists although they do, like the undergraduate physics community of practice, share legitimate peripheral practices. The person he talked to informed Larry that if he wants to finish his physics degree, his best approach to becoming a teacher will be to go to graduate school and take education classes that will equip him to become a teacher.

*L3.4 Exploration.*

Larry: Like the people I talked to said now is not the right time to apply since I am not getting a teaching certificate now. But I'm a promising candidate and so once I go to grad school, I will just get a master's in physics and a teaching certificate at the same time and they said then that would be the time to apply. So, financially that's kind of my plan and then, so I have thought about it at this point.

The caveat to this increase in exploration of his identity and future career is that he still displays an uncertainty in his commitment to teaching despite all of his positive affect towards the profession.

*L3.5 Future plans.*

Larry: You know, I like physics or I wouldn't have stuck with it… so more physics in grad school doesn't exactly sound bad.

Int: Okay.

Larry: And then in case I don't like being a teacher, I have a master's in physics so that's a little bit of a cushion there.

Larry's worry is perfectly understandable but displays a lack of comprehensive exploration on his behalf. Larry had the opportunity during the last year to explore teaching through practical experience by getting involved in the university's teaching assistant program. Several of Larry's peers have taken advantage of this program and have been engaged in teaching for some time. Larry also has had the opportunity to join a research group and explore his perceived dislike for that aspect of physics. Up to this point his commitments to band (marching band) have prevented him from pursuing either but he maintains (as he has previously) that his priorities need to change, and he needs to pursue one of those options in the coming year.

*L.3.6 Choices/crisis point.*

Int: It sounds like a choice there. You can investigate teaching through a TA position or join a research group and figure out what research is all about.

Larry: Hmm, yeah, and I have thought of that and I'm really not sure which one to go with.

At this point in Larry's undergraduate career he still has a lot of exploration to conduct before moving into a *moratorium* identity status with his professional identity and some big decisions to make before he will get to a place where he has an *achieved identity* status. By the end of the third interview, he has started displaying some progress by actively exploring how he might achieve his intended goal of teaching. Larry has demonstrated through his constant positive affect towards physics and his continued commitment that he is still in a *foreclosed* subject-specific identity. Larry has not though explored any of the other options available to him as he pursues his physics degree.

|                                               |                                           |
| --------------------------------------------- | ----------------------------------------- |
| Identity Achieved<br><br>High Commitment<br>High Exploration | Moratorium<br><br>Low Commitment<br>High Exploration |
| Foreclosed Identity<br><br>High Commitment<br>Low Exploration<br><br>[Int 1] [Int 2] [Int 3] | Identity Diffusion<br><br>Low Commitment<br>Low Exploration |

**Figure 8**: Larry's identity status in regards subject-specific identity with each interview.

|                                               |                                           |
| --------------------------------------------- | ----------------------------------------- |
| Identity Achieved<br><br>High Commitment<br>High Exploration | Moratorium<br><br>Low Commitment<br>High Exploration |
| Foreclosed Identity<br><br>High Commitment<br>Low Exploration | Identity Diffusion<br><br>Low Commitment<br>Low Exploration<br><br>[Int 1] [Int 2] [Int 3] |

**Figure 9**: Larry's identity status in regards personal identity with each interview.

Over the span of the three interviews, Larry has not displayed much change or progress in either his professional or subject-specific identity. The lack in change in his professional identity could be associated with his perception of himself as already being a physicist. This perception plus his rejection of practices that are more central to the community of practicing physicists has resulted in very little exploration on his behalf and hence being stuck in a *diffusion* identity status for his professional identity as illustrated in figure 9. The amount of time left in his undergraduate degree may force an identity crisis but his decision to pursue graduate school to allow him the option of becoming a teacher may further postpone his pursuit of authentic experiences and making a decision about his future with physics. It may afford him the opportunity to engage in more central legitimate peripheral practices of the community of practicing physicists though. In regards his subject-specific identity, Larry, has demonstrated a high level of commitment throughout all the interviews. He is definitely committed to obtaining a degree in the subject of physics. We have labeled Larry as being foreclosed as opposed to achieved in regards the status of his subject-specific identity for the period of the interviews as indicated in figure 8. We decided foreclosed as opposed to achieved for two reasons. The first is that Larry displays a normative identity processing style that is evidenced by his appropriation of goals of important others like Neil deGrasse Tyson. Larry also does not indicate that he has explored the physics topics available for him to study or the opportunities that are available to him in other subjects. This lack of exploration indicates a foreclosed identity.

**Case 3: Bob**

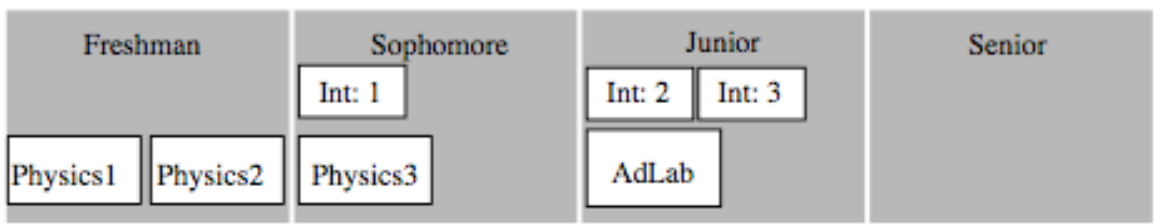

**Figure 10**: Data collection timeline for Bob. Physics 1 and 2 refer to introductory physics 1 and 2 which cover mechanics and electricity and magnetism. Int 1, 2 and 3 refer to the different interviews conducted with the student. AdLab refers to the advanced laboratory class.

Bob was 19 and a sophomore when first interviewed. His first experiences with physics were in high school, in his junior year. He describes it as neither a positive nor negative experience. He took physics as a senior in high school because he was good at math. Although unsure of what he wanted to do, he had the idea that he wanted to be an engineer. His experiences of physics as a senior were similar to that of a junior, neither positive nor negative. He decided to come to university to major in mechanical engineering but in the summer prior to starting he became more interested in physics and started to read research papers in several areas of physics. At the time of the first interview, Bob is in Physics3 and has taken Physics1 and Physics2 the previous year.

Interview 1

In the first interview, Bob is a sophomore and describes adding physics as a major at the end of his freshman year. At the time of the first interview, Bob's decision to add physics as a major is a very recent identity crisis event for Bob.

*B1.1 Decision to take physics.*

He describes a time of reflection leading up to this decision where he was in the process of becoming more and more interested in physics.

Bob: <u>I guess</u> the actual science, instead of the engineering. <u>I still like</u> the engineering as well. I just decided to double major in them. That's why I'm in… I didn't declare physics until like the start of my last year, my freshman year, but it was… <u>I had pretty much</u> planned on it. I initially came here for engineering, but <u>I pretty much just decided</u> that I wanted to do physics as well and that's probably what I am most interested in.

Bob's exploration process involved reading physics articles for fun while he was taking Physics1 and Physics2. This exploration of his interest in the subject helped him decide to pursue physics as a major. Bob displays similarities to Sally in their decisions to pursue multiple majors. Sally makes several comparisons between chemistry and physics; Bob makes comparisons between engineering and physics. These comparisons are a good insight into the exploration and reflection process Bob is engaged in as he distinguishes differences between the subjects and why he is currently more inclined towards pursuing physics as his primary identity rather than engineering.

At this point in Bob's pathway he makes two distinctions between physics and engineering as illustrated in the sections below. The first is:

*B1.2 Difference between engineering and physics.*

Bob: Engineering is just applying stuff and physics is like… you have the chance to see things that no one else has ever found.

Later on in the interview Bob continues talking about the difference.

Bob: Engineering is kind of easier and the main reason that is, is because a lot of the students going into it, there's a lot more of them and there is a less percentage of smarter students, it seems like, like in physics. I feel like if you're a physics major, like lots of people who go into engineering, they see the money, you know… they are kind of doing it for something like that. I feel like if you're going into physics, you're more interested in the topic than people going into engineering really… really aren't that interested in it. They were maybe decent at math in high school and saw the dollar signs. You know, whatever… I'll do engineering, that makes the classes easier cause when they are [curving] it, you only have to beat the other students, so the engineering classes are definitely easier.

It would seem to Bob that physics is a grander enterprise than engineering and this idea comes from the perception that physics is more difficult and is a more unique major. In the affective domain, Bob emphasizes an interest in physics over engineering but his main affective response to physics seems to be less emphatic when compared to Sally and Larry. They both expressed strong emotions about particular aspects of the subject. Of course, this could be a result of Bob's personality, as opposed to an indication of his lack of commitment to physics. However, in the AdLab observation data, Bob becomes enthusiastic during a number of off topic discussions with other groups and displays an excitability for certain topics akin to that of Sally and Larry during their discussions on physics.

Bob does not yet consider himself a physicist. As a result of this perception, he does not share Larry's ethos that if he has the mindset of a physicist then he is a physicist. However, Bob also does not share Sally's idea of commitment to the subject as being the decider as to whether one is a physicist. Instead, Bob believes that one becomes a physicist when they have received a qualification such as Ph.D. or degree in physics.

*B1.3 Perceptions of what it means to be a physicist.*

Bob: Well I suppose it depends on the definition, but I guess most of them will have Ph.D.'s in physics that would qualify being a physicist.

Bob is also distinctive from the other two students in that he has a very specific career in mind: Bob wants to work for NASA, and he talks about this with much more emphasis than his interest in physics.

*B1.4 Future plans.*

Bob: One of my <u>passions</u> is that I want to work for NASA as either a scientist or an engineer, so if I have a degree in physics and engineering it can't hurt.

It is obvious though that Bob is still unsure about the specifics of this path to working for NASA. He tends to communicate with hedging (see: B1.1) when speaking about the specifics of his choices so far: "I guess the actual science"; "I pretty much just decided that I wanted to do physics as well, and that's probably what I am most interested in." Larry did not display hedging because he is not distinguishing between two majors and Sally is much more definitive about her preference for physics from the start, but does do some hedging until the final interview. Hedging is an indication of being in either a *moratorium* identity status or an *identity diffusion* identity status. Bob makes it easier to classify him as being in a *moratorium* identity status for both of his professional and subject-specific identities as he is exploring both the subjects of engineering and physics and he is looking to join a research group. The hedging that he presents when talking about a preference for either subject displays a lack of commitment to either of his subject-specific identities (engineering and physics).

At the time of the interview he is "debating between a few [research fields]. You know, I don't really know that much about it yet. I'll take the classes and probably decide, but thinking about some kind of high energy physics… you know like particle physics and stuff. I was kind of leaning towards cosmology stuff too, like astro, but I don't really know yet honestly." Again he displays an uncertainty in his willingness to commit to a specific area but he intends to make a decision based on his experiences in his classes. This is a form of exploration and self-reflection that fits with an informational identity processing style that in turn indicates a *moratorium* identity status.

Interview 2

As with Larry, the second interview with Bob took place six months after the initial interview and it is also the start of his first semester as a junior. One of the classes he is taking this semester is AdLab. The similarities continue between Bob and Larry as Bob's identity status remains the same and is still in a *moratorium* physics identity status in relation to his subject-specific and professional identity. The evidence of his *moratorium* status is in his statements of still being unsure about which of his majors to pursue in the future, and he again uses hedging words when describing the possibilities that lie ahead.

*B2.1 Future plans.*

Int:  Why are you double majoring as opposed to concentrating on one?

Bob:  Uh… because… uh… <u>I don't know</u>. I liked both of them. I <u>wasn't really sure</u> what I wanted to do…. still not, so this gives me <u>more options.</u>

Int:  Are you going to continue with the double major or let one go to a minor

Bob:  I'll <u>probably</u> continue with it because um… um…um… like engineering, you can't really get a minor in, um physics. I'm <u>pretty sure</u> I want to get a major in that so...

Bob discusses his uncertainty directly by stating his uncertainty about the future but also indirectly by his use of hedging words when it comes to discussing his plans. Bob continues to make comparisons between his two majors again indicating that he feels engineering is an easier subject, but also that some of the engineering courses are just different versions of some of the physics classes that he has taken.

*B2.2 Similarities between engineering and physics.*

Bob:  The engineering class I'm in is dynamics, which is pretty similar to mechanics I think, but easier.

Int:  Why is it easier?

Bob:  Because it's the engineering version.

Int:   Do you find that a lot, the engineering versions?

Bob:  Well, just the engineering classes in general.

This similarity between engineering and physics will become important in Bob's next interview. Also important is his distinction that engineering is easy and so by association physics is hard. For Bob, the fact that physics is hard is attractive. Bob's affective response to physics continues in that he finds it more interesting and again makes unflattering comparisons to engineering.

*B2.3 Positive affect.*

Bob:   Yeah, I'd say I'm leaning more towards physics.

Int:   And why do you think that?

Bob:   'Cause it's more interesting. And lots of engineering jobs are boring and you sit at a desk with a whole bunch of engineers and… I don't know, physics is a lot more… I don't really know how to put it. I guess a lot less people do it and there is just more interesting stuff.

Bob again indicates in the interview that he is still looking to obtain a research experience and that he still considers a Ph.D. and research essential to being considered a physicist.

Observations

In the case of Bob, as opposed to Larry and Sally, he does not stay enrolled in AdLab. After three weeks into his AdLab semester, he drops out of his physics major. Bob drops out before his third interview, and the researchers arranged a final interview to discuss his decision. The observation data is particularly applicable to Bob as his interactions indicate uncomfortableness in the AdLab physics learning environment, which may be related to him being on the verge of an identity crisis incident and his decision to drop physics as a major. During the two-week period that Bob was enrolled in the AdLab environment, he makes several affective statements regarding his self-efficacy in this environment. It is important to note for the two-week period that Bob was a member of the lab, he was also a member of Larry's group.

Bob and his group are setting up an experiment involving a complicated circuit, which is illustrated through a circuit diagram. The components have to be sourced from around the room and the power source is particularly complicated to set-up correctly. At the time of the quote Bob and his group have been setting up the experiment for about 20 minutes and Bob has been talking to his group members but being standoffish with the equipment and setting-up anything.

*B0.1 Negative affect.*

Bob:   I am not very good at reading these diagrams (referring to circuit diagrams).

A similar comment occurs again later on after the instructor has helped the group set-up the experiment:

Toby:  (directed at Bob and pointing to the experiment setup) There are so many different parts to that experiment.

Bob: (looks at the setup and shakes his head up and down in agreement) I have no idea what is going on.

(Toby laughs)

(Bob pauses and looks into the distance)

Bob: I'm really not good at it. She (the instructor) set that up.

In another conversation that Bob initiates about a class that the group all enrolled in that is running concurrently to the AdLab course, we get another insight to Bob's struggle with his self-efficacy in physics.

*B0.2 Negative affect.*

Bob: Are you in mechanics?

(Larry shakes his head yes.)

Bob: What do you think of it?

Larry: Homework is hard.

Bob: I haven't really started it, but I looked at it and it looks confusing…It's just like, I feel like it's stuff you never really practice. You know what I mean?

Larry: We're just diving in head first (laughs and makes distance gestures with his hand). I think the bottom of the pool is like two feet away.

(Bob nods in agreement but does not seem to find what Larry just said as humorous.)

Bob: It like, makes sense, but doesn't come really fast.

Bob is evidencing much negative self-efficacy when it comes to his self-concept of his abilities in physics during this period. There is a similar theme between Larry and Bob in regards to how they are engaging in legitimate peripheral practices of the community of practicing physicists but for different reasons. Larry does not want to engage in more central practices of the community of practicing physicists, as he does not believe that he will enjoy it as much as he does engaging in the practices of teaching and solving problems. Bob on the other hand does not want to engage in the more central activities of the community of practicing physicists, as he does not believe that he will be good at them. Setting up and designing experiments is one of the most central legitimate peripheral practices that an undergraduate student can engage in the community of practicing physicists and Bob does not think that he is good at it. Another germane conversation related to the themes discussed is as follows.

*B0.3 Exploration.*

Bob: What are you doing in the basement?

Matt: Right now we are setting up to spectroscopy using rubidium.

Bob: How long do you… how many hours a week is it?

Matt: Like four or five.

Bob: Yeah. I've been wanting to get into something, but I don't know when I can do it, 'cause don't they usually work 8 to 5 or something

Matt: I'm just in there Tuesday and Thursday mornings, 'cause that's when I have free time. So it really just depends on when the professor is free.

Bob: Yeah. I just think it would be really hard to find the time when I'm free and the professor is free.

Matt: Hmmm.

Bob: 'Cause I have stuff going on between 8:30 till 4:30 every day. I don't think an hour counts, so it/s like, I don't know when I'd do it.

Matt: Well you could talk to them and see if you could come in for an hour at a time, like three or four times a week. And a problem is, a lot of the time, I'm down there, I'm by myself, so I just talk to [my research professor] Dr. Bridges, 10, 15 minutes when I get in, see what he wants me to do. You'de probably be able to find someone who can work with your schedule.

Bob: Can you just go in and ask?

Matt: Yeah.

Bob: Did you even know him before?

Matt: He gave a talk in Physics Today [an introductory seminar for physics majors] and I went up to him after and asked him if I could work with him and he was like, 'Yeah'.

Bob: Yeah, 'cause I really do need to start getting some experience… maybe one hour every once and a while.

Matt: Yeah. You should be able to find someone who you can do that with.

The previous discussion is very interesting as Bob is still exploring the idea of gaining research experience, but it would seem as though he would have difficulty fitting it into his current schedule. There are also signs that this frustrates him, as he knows he has to do it.

Another important note about this conversation is that although it is an interaction between Bob and Matt, Larry is, in fact, present during the conversation and does not join in even though he has indicated his own need to pursue a research experience. Bob would appear to still value it as an important experience but only if it were able to fit his schedule.

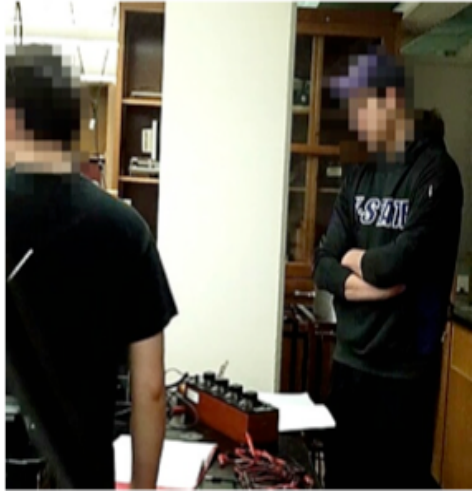

**Figure 11**: Bob (right of picture) arms folded and standing back, reluctant to get involved with the experimental equipment.

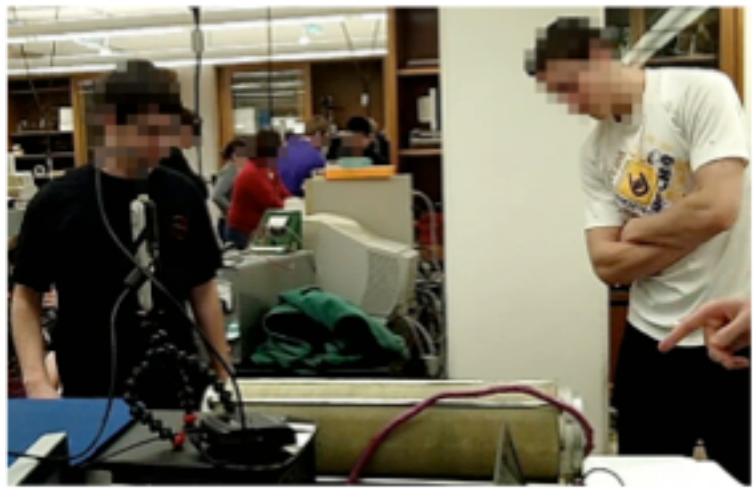

**Figure 12**: Bob's (right of picture) again with arms folded and standing back, reluctant to get involved with the experimental equipment.

Finally another point to note that was identified in the observational data was Bob's body language during the laboratory sessions when compared to both Matt and Larry. Bob consistently had his hands in his pockets or folded and rarely involved himself in the actual experimental process as evidenced in figures 11 and 12. This body language is a clear indicator that he is not comfortable in the learning environment nor with the activity he is engaged.

Interview 3

At this point in Bob's academic career he has decided to drop physics as a major, describing it as a practical decision, as he felt the two subjects did not complement each other. Bob describes how his identity crisis manifested after some reflection, and it would seem as though for now he has entered into an identity-achieved status with his subject-specific engineering identity. Bob has previously described the distinction between physics and engineering as the engineering being easier. Up to this point, this distinction was a cause for justification for Bob doing physics as a major and he seemed to be quite dismissive of engineering.

*B3.1 Crisis point.*

Int:   So you decided to finish with your physics major? Why was that the case?

Bob:   I started thinking about it more and I still wasn't sure which one I wanted to do and I realized I had three more years of really hard semesters and not having the other one benefit the other one would kind of be like a waste, you know. Like doing two majors, doing three more hard years and no matter which one I choose the other doesn't really benefit the other one that much. I was thinking about it and I thought it would be easier to make a choice. It would make my life a lot easier.

The statement that one does not benefit the other may be a result of Bob's conceptions of physics and engineering. In a simplistic sense, Bob sees engineering as an easier version of the subjects being covered in physics. As he indicated before, you cannot take engineering as a minor and with this conception, engineering is not helping with his physics, so he might as well drop the physics. Bob's moratorium identity status in regards to which subject-specific identity to commit more to seems to be part of his reasoning to make a more definitive choice. Another element of this crisis point arising was that Bob had continued to try and seek a research experience. He had applied for several REU's (Research Experience for Undergraduates) and was not optimistic of his chances and during this application process he was offered a summer position with Boeing.

*B3.2 Crisis point.*

Bob:   I had the opportunity to get an internship with Boeing because I want to go into aerospace and… and do astrological engineering. And I found an opportunity for that and ended up getting that.

In the end, Bob's exploration of both identities ended up at crisis point. The resolution of this crisis being resolved was to choose whichever subject community was easier (or perceived easier) to become a more central member as soon as possible. Bob was aware that he would need to engage in more central legitimate peripheral practices of one of his possible professional identities in order to become a more central member of that community. It turned out that it was easier to gain that experience with more central practices in the community of practicing engineers than in physics. It also didn't hurt that the subject-specific identity that is related to this professional identity is perceived to also be easier. The interesting point about Bob's choice is that he is not discounting becoming a member of the community of practicing physicists in the future.

*B3.3 Future plans.*

Bob:    I like science in general more. But if I get a master's in engineering, what I want to do is fluids or aerodynamics, which is pretty physics intense anyways. So I look at it as kind of like… I still pretty much do both things. I like… Anyway, that's what I want to do anyway… is advanced aerodynamics or fluids… something like that, that I can apply to spaceflight. Something along those lines…. In reality I will be using physics all the time, but I can just get an engineering degree and get a master's and something in engineering will be mainly physics, but just not the name on it, I guess.

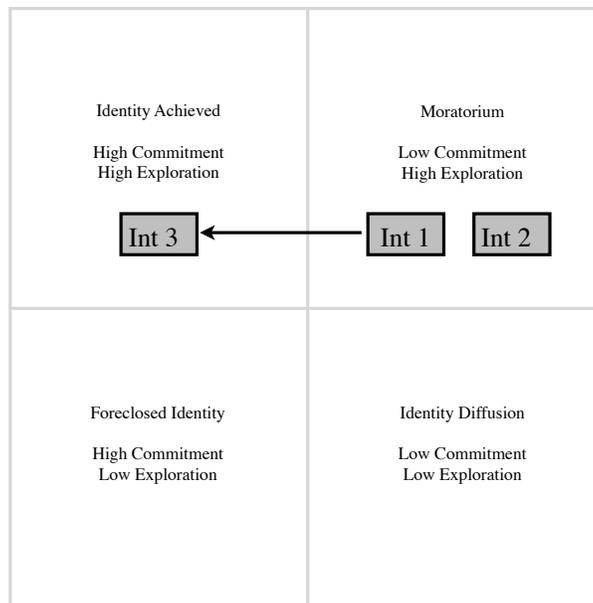

**Figure 13**: Bob's identity status for his subject-specific identity with each interview.

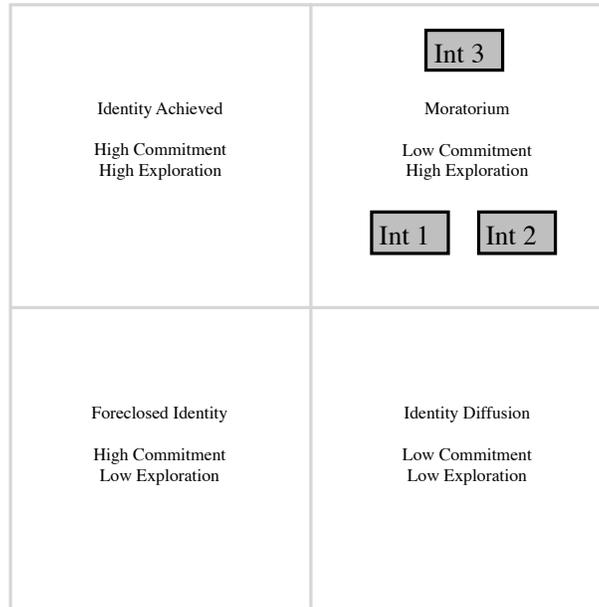

**Figure 14**: Bob's identity status for his professional identity with each interview.

In essence Bob believes that he can still be a member of the community of practicing physicists later in life and still engage in some of the more central practices of the community of practicing physicists because of the overlap between his intended occupation and the community of practicing physicists. An occupation however is not a professional identity and Bob's statement in B3.3 would indicate he still has a *moratorium* identity status in regards his professional identity. His decision to focus on engineering has resulted in a change from moratorium identity status to an identity achieved status for his subject-specific identity as illustrated in figure 13 but his professional identity remains in low commitment (Figure 14) and hence he is still in a moratorium identity status.

**Discussion**

Identity status and communities of practice

The three case studies presented indicate an argument for a relationship between communities of practice and identity statuses. The relationship between these two constructs needs to be investigated further, but it can be argued that a positive experience when engaging in practices that are perceived to be more legitimate and more central to a professional identity can result in a change in identity status. This is evidenced by Sally's experiences as an undergraduate researcher, which have resulted in her visible transition to an *identity achieved* status and being a more central member of her research group and the community of practicing physicists. A negative experience with more legitimate and more central practices of a professional identity can result in no change in professional identity status. Larry and Bob's experiences in the AdLab learning environment both had negative aspects and resulted in no change in their professional identity status. The positive relationship makes sense, as students who begin to become more core members will explore more specifically the practices of the community they are joining.

Students are then better able to assess whether it suits their goals and whether they have sufficient positive affect for the subject. The negative relationship highlights a possible problem with the typical pathway for a physics student to becoming a physicist.

In the past there have been pathways to becoming a physicist that did not involve obtaining a physics degree but that pathway is not the typical approach to central membership of the community of practicing physicists. The practices and understanding that students engage in, in their undergraduate physics career, are legitimate peripheral practices of the community of practicing physicists. At some point in their undergraduate or graduate career, students begin to break through the boundary conditions of the legitimate peripheral practices of the undergraduate community of practice and begin to engage in more central practices of the community of practicing physicists. At this point they move from central membership in their subject-specific identity to a trajectory for more central membership in the community of practicing physicists. In the case of the three students presented in this paper this breaking of a boundary condition occurred for only one of them: Sally. This research indicates that Sally's involvement with research and by extension more central and legitimate peripheral practices of the community or practicing physicists has resulted in a change in her identity status.

One could make the argument that if the AdLab learning environment had provided a more authentic experience of conducting research and engaging Bob and Larry in legitimate peripheral practices of the community of practicing physicists then it might have helped change their professional identity status to identity achieved in relation to physics. But Larry and Bob's cases do not provide enough evidence to make this claim and in fact offer alternative interpretations. Bob's story would seem to indicate that a level of preparedness is needed in order for him to believe in himself enough to not feel unable to participate in the AdLab learning environment. This could be an argument of communities or practice gone wrong where a student has been guided towards a more central place in the community before he or she is ready. Larry on the other hand opens up a question about the construct of community of practice and its application to the college setting. In the community of practice a person moves into the center of the community by engaging in legitimate peripheral practices of the community guided by central members. In a college course setting, especially in a science subject, the actual legitimate central practices of the community often cannot not be engaged in, until after you have obtained your degree. Therefore the practices and understanding that the professor (professional physicist) is guiding you through are often legitimate but somewhat peripheral practices of the community of practicing physicists. So when Larry perceives himself as being a physicist and that research is not necessary to become one this is an understandable interpretation given the practices that he is being asked to engage in. This means that when applying the community of practice framework you cannot just assume that central members are effectively communicating what the legitimate central practices are of that community. If they are not being effectively communicated then peripheral members of that community may not feel the need to seek out new, more central, and legitimate ways of participating in the community and so will not be on a central trajectory. They will end up believing that they are participating as central members and their identity development will become stagnant.

Relationship to previous identity status research

According to Michael Berzonsky et al. (1990), more advanced statuses are orientated toward the future. Verification of this can be found in one of the students (Sally) in this study who had a specific long-term goal that was orientated towards the future. Finding a specific career for which to strive for is an argument for opening up more opportunities or activities that ask students to explore the possibilities open to them. Another argument made in the past is that those in an *identity achieved* status are more likely to self-assess as being a part of their professional community (Côté and Levine 2002). In our study, this point would also be applicable to Sally. This relationship once again reveals a possible relationship between communities of practice and identity status as Sally who explored her relationship with her subject-specific identity and professional identity has engaged in the legitimate practices of the community and has thus become more accepted.

In the domain of affect, previous research has illustrated that low-commitment statuses (e.g. *moratorium*) have found to be correlated with higher anxiety and more negative self-beliefs (Luyckx et al. 2008). Bob would seem to fall into this category with his physics identity, as in class he displays doubts about his self-efficacy in the AdLab environment and shows anxiety about obtaining an appropriate research experience in physics. Larry reinforces this finding also as his committed *foreclosed* subject-specific identity status results in high self-belief in physics and a lack of anxiety displayed in his interviews and observations of him working in the laboratory.

Manifestation of identity status in students' affective responses

The case studies presented in this paper show that students' affective responses can be indicative of their relationship with their subject-specific identity. To hear Sally talk in the AdLab environment, one could not be confused as to her commitment to her subject-specific identity as a physicist. On the other end of the spectrum Bob's displays of negative body language and reflections on his lack of ability in regards to particular physics-related activities would be indicative of a troubled relationship with his physics subject-specific identity. The observational evidence presented is indicative that a further investigation of the relationship between identity statuses and affective responses in the classroom might be a worthwhile course of study and could result in the obtaining of a rubric that could help professors identify students who are having difficulties with their identity and hence their commitment to the subject.

Appropriation of 'professional physicist' practices

Hasse and Trentemøller (2008) argued that the physics culture favors an interest for research and the subject itself rather than applications. Larry's perceptions do not fit this description in that his view of research not being important and his preference for problem-solving would appear to put him out of sync with the culture of the subject in which he has his subject-specific identity. An argument to understand this discrepancy is that many university physics programs do not cater to Larry and his ilk as universities are trying to create more physicists in the mold of those that they already have (professors, the majority of who conduct research). Larry only wants to enculturate some aspects of the practices of the community of practicing physicists and ignores others and he essentially needs guidance as to what else a physicist can do besides research. He especially needs guidance in relation to which practices are important for him to appropriate for his proposed future role as a high school teacher. Physics departments may be inadvertently sending the message that future physics teachers will not be physicists due to the focus on appropriation of research-orientated practices. The absence of a community of practice of physics teachers within the degree track also limits his ability to engage in the legitimate peripheral practices of that community. But it must be pointed out there is a potential community of practice of undergraduate learning assistants that Larry has not tried to gain membership to as of yet.

Undergraduate intervention

The students in these three case studies take different pathways through physics and value different aspects of the practice of physics. It would seem that all three of the students in this study would have benefitted from more guidance and possible interventions at key moments to help them process their identity development and current/future identity crises. With this help the students could focus on their desired professional identity and its related end goal relatively more comfortably when they would not have important life decisions weighing on them. We are not saying that one should rush students into making this important decision before an adequate amount of exploration has taken place. However, some students might need an educational intervention that fosters both metacognitive/epistemological sophistication development and provides students with opportunities to experience more core elements of the practices of the community they intend to join. As discussed earlier, the informational processing style associated with the *moratorium* and *identity achieved* statuses may be developed by encouraging students to develop meta-cognitively. By developing students' processing styles, we may better equip them to handle the difficult situations that await them at crisis points.

It is clear from Sally's case that her research experience had a significant influence on her pursuit of physics as a major and professional identity. But in the cases of Larry and Bob, neither attains the requisite research experience. Bob seems to base some of his decision-making on which major he can get a relative experience in while Larry still intends to pursue a research experience, but without displaying the drive to actively obtain one.

The undergraduate research experience is an example of an educational intervention and can be a powerful tool for identity development. Students are substantially encouraged to pursue such an experience at Kansas State University. It may not be adequate by itself though, especially when students do not gauge the importance of such an experience to their professional development. Larry is a clear example of this in that he acknowledges that a research experience is encouraged; yet he does not actively pursue one. Maybe he does not feel it is essential. He may not be epistemologically sophisticated yet to understand the importance of exploring one's possibilities in a substantial manner. Bob is similar at the beginning of the interview process in that he puts off gaining research experience until it fits his schedule, but leading up to his identity crisis point he seeks relative experiences in both physics and engineering.

In previous research, Tobias Krettenauer (2005) found that two identity statuses, *moratorium* and *identity achieved*, partly depend on the development of epistemic sophistication. Jane Kroger and colleagues (2010) also indicate that a student's time in university will be associated with more transitions through *moratorium* status than other ages. With the prevalence of the *moratorium* status, we must provide students with some opportunity to develop their epistemological sophistication. Previous research (Kroger and Marcia, 2011) has indicated that identity-based interviews can help students in a *moratorium* status figure out their identity and obtain an *identity achieved* status. There is evidence in this study that Sally used the identity interviews to help attain her *identity achieved* status.

The implications of Sally's journey indicate that there are aspects of both the undergraduate research experience and the identity interviews that provide assistance in progressing from the

*moratorium* status. Future studies should focus on discovering what aspects of both of these experiences lead to positive identity exploration of a professional and subject-specific identity and ascertain whether these aspects can be replicated in creating learning environments that offer a similar opportunity. It is difficult to promote substantial identity development through short-term intervention programs. A program encouraging identity reflection and exploration whilst also addressing epistemological sophistication should be implemented concurrently with students' subject-specific studies. Resolving these issues in the undergraduate domain is important as an *identity achieved* status when graduating from university will transfer to the workplace (Luyckx et al. 2010).

**Conclusion**

This study showed three different pathways through physics with three different students. Each student showed different transitions through distinctive identity statuses and enabled correlations to be made between identity statuses and communities of practice and affect and identity statuses. The paper highlights the importance of undergraduate research as a method of identity exploration but also the substantial need for other educational interventions that result in students reflecting and exploring their identities.